\documentstyle[twocolumn,seceq]{jpsj}

\def\runtitle{Ferromagnetism in Lieb-Ferrimagnetic Models}
\def\runauthor{Yuuki {\sc Watanabe} and Seiji {\sc Miyashita}}

\title
{
Existence of Saturated Ferromagnetic and Spiral States \\
in 1D Lieb-Ferrimagnetic Models away from Half-Filling
}

\author
{ 
Yuuki {\sc Watanabe}\footnote{E-mail address: 
wata@godzilla.phys.sci.osaka-u.ac.jp}
and Seiji {\sc Miyashita}\footnote{E-mail address: 
miya@ess.sci.osaka-u.ac.jp}
}

\inst
{
Department of Earth and Space Science, Graduate School of Science, 
Osaka University, Toyonaka, 560-0043
}

\recdate
{
March 15, 1999
}

\abst
{In order to study conditions for the appearance of
ferromagnetism in a wide filling region, 
we investigate numerically three types of 
one-dimensional Lieb-ferrimagnetic Hubbard models: 
a periodic diamond (PD) chain, a periodic alternately-attached leg 
(PAAL) chain and an open diamond (OD) chain. 
All of these models have a flat band (or equivalently, degenerate 
single-electron eigenvalues). The PD and OD chains 
commonly have a local-loop structure. Nagaoka's theorem holds only 
in the PD chain. 
At half-filling, it have been rigorously proven that 
all of these models are ferrimagnet.
Away from half-filling, however, quite different magnetic properties 
are found. 
In the fillings $1/3<\rho_{\rm e}<1/2$, the ground state of 
the PD chain for a infinitely-large $U$ is the extended ferromagnetic 
state, that is, the saturated ferromagnetic 
state or the spiral state for odd or even number of electrons, 
respectively. 
In the PAAL chain, on the other hand, there is no 
magnetic order. 
Thus, the flat band is found to be not a sufficient condition 
of the extended ferromagnetic state.
We find, moreover, that the saturated ferromagnetism appears 
in the OD chain, 
although the Nagaoka theorem does not hold on this chain. 
This indicates that the local-loop structure plays an important role 
on the appearance of the extended ferromagnetic state. 
}

\kword
{
Hubbard Model, Flat Band,  Nagaoka's Theorem, 
Saturated Ferromagnetism, Spiral State, Ferromagnetic State in 
an Extended Sense, Numerical Diagonalization
}

\begin{document}
\sloppy
\maketitle

\section{Introduction}
 Ferromagnetism in strongly correlated electron systems has 
attracted great interests. Since ferromagnetism is a purely 
many-body phenomenon and appears for sufficiently-strong  
interactions and high electron fillings in general,~\cite{TH0,PDBDF1} 
we have to investigate properties of the ferromagnetism by 
non-perturbative methods.
The important question whether the Coulomb interaction 
is enough to realize ferromagnetism is still far from completely 
understanding. 

 Up to recent, two rigorous results concerning with the 
existence of saturated ferromagnetism have been known on 
the Hubbard-type models,\cite{GMC1,HJ1,KJ1} which consist of 
the hopping term and the on-site Coulomb interaction term.
They are Nagaoka's theorem,\cite{NY1,TDJ1,TH1} and 
(nearly-)flat-band ferromagnetism.\cite{MA1,TH2,MT1,TH3}

 Nagaoka's theorem holds only for the system with just one hole 
added to the half-filling. 
Therefore, after Nagaoka's theorem was found, it 
has been repeatedly studied from various points of view 
whether Nagaoka's theorem holds or not for more than one hole. 
In the two dimensional square lattice, 
many authors show that the saturated 
ferromagnetic state is no longer the ground state for 
the two-hole case.~\cite{TM1,DW1,RY1,FRDSR1} 
Kusakabe and Aoki, however, showed that the ground state of the 
two-hole system is not the ordinary singlet state but 
the spiral state.~\cite{KA1} The definition of the spiral state 
is that its spin structure factor $S(Q)$ 
takes its peak at $Q=2\pi/L$ ($L$: system size).

 On the contrary, the flat-band ferromagnetism is realized at the 
half-filling of the lowest-energy flat band, for example, the 
quarter-filling in the one-dimensional Mielke-Tasaki model. 

 Between the one hole added to the half-filling (Nagaoka's 
ferromagnetism) and the quarter-filling (flat-band ferromagnetism), 
a new type of ferromagnetism has been found 
in a one-dimensional Mielke-Tasaki model.~\cite{WM1} 
The ground state in this filling region has been shown to be 
the saturated ferromagnetic state or the spiral state corresponding 
to the odd or even number of holes, respectively.~\cite{WM2}
The spins in the spiral state have a ferromagnetic 
correlation up to half of the system size. Therefore, we regard both 
states, the saturated ferromagnetic state and the spiral 
state, as a ferromagnetic state in an extended sense. 
Hereafter, we call them the extended ferromagnetic state.

 In this paper, to study conditions to realize the extended 
ferromagnetic state in high electron fillings, 
we investigate three categories of models, (A), 
(B) and (C). To classify the these categories, we take notice 
of the following three characteristics: existence of the flat band 
(or equivalently, existence of the degenerate eigenvalues 
in the single-electron state), existence of local loops (triangles 
or squares) and 
holding of Nagaoka's theorem. Although the existence of the flat band 
is a common characteristic in the categories (A), (B) and (C), 
the local-loop structure exists in (A) and (C), and Nagaoka's 
theorem holds only for (A). Here, we note that all the three 
characteristics are satisfied in the flat-band 
ferromagnetic models which was previously studied. 
In this paper, therefore, we would treat other Hubbard-type models. 

 First, we compare two models of (A) and (B). 
As we will see in \S 2, at high (from more than $1/3$ to less than $1/2$) 
electron fillings, we find that the extended ferromagnetism  
(not ferrimagnetism) appears in a model of (A) while all the ground states 
at less than half-filling are singlet or (disconnected) paramagnet 
in a model of (B). Thus, we conclude that the existence of the flat band 
is not a sufficient condition for the appearance of 
the extended ferromagnetic state. 
Furthermore, we find that in a model of (C), the saturated ferromagnetic 
state appears at high electron fillings in spite that the model dose not 
satisfy the condition for Nagaoka's theorem to hold. This indicates 
that the existence of the local-loop 
structure plays an important role for the appearance of the 
extended ferromagnetic state. 
We would stress that this is the first result about the appearance 
of the saturated ferromagnetic state in such a wide range of fillings 
on a system to which Nagaoka's theorem is not applicable. 

 The rest of this paper is organized as following. In the next section, 
models and some definitions are presented. Rigorous results related 
to the present problem are also reviewed. 
In \S 3, numerical results for models of (A) and of (B) 
are presented. 
The result for model of (C) is obtained in \S 4, 
along with the discussion about the relation between the local-loop 
structure and the connectivity condition. 
Summary and further discussions are given in \S 5.

\section{Models and Related Statements}

 First, we define the models and some basic definitions 
treated in this paper. 
The Hamiltonian on a lattice $\Lambda$ is 
\begin{eqnarray}
 H&=&H_{\rm hop}+H_{\rm int} \\ \nonumber
  &=&-t\sum_{\langle i,j \rangle \in\Lambda}   
  \sum_{\sigma=\uparrow,\downarrow}
    c^{\dagger}_{i,\sigma}c_{j,\sigma}
  +U\sum_{i\in\Lambda}
  n_{i,\uparrow}n_{i,\downarrow},
\end{eqnarray}
\noindent
where $U>0$ is the on-site Coulomb repulsion energy, $\langle i,j\rangle$ 
denotes the nearest-neighbor pair in the lattice $\Lambda$ and 
$c^{\dagger}_{i,\sigma},$ $c_{i,\sigma}$ and $n_{i,\sigma}$ 
are the creation, the annihilation and the number operators of the 
electron on site $i$ with spin $\sigma$, respectively. We assign 
the hopping term $-t <0$ to each bond. 
$N_{\rm e}$, $N_{\rm h}$ and $\rho_{\rm e}$ denote the number 
of electrons, the number of holes and the electron filling, respectively. 
Here, $N_{\rm h}=0$ when the whole system is half-filled. 
We consider the following models as examples for the categories, 
model(A), model(B) and model(C):
\begin{description}
 \item[(A)] PD chain: diamond chain with the periodic boundary condition.
 \item[(B)] PAAL chain: alternately-attached leg chain with the periodic 
			boundary condition.
 \item[(C)] OD chain: diamond chain with the open boundary condition.
\end{description}
\noindent
The diamond chain and the alternately-attached leg chain are shown in  
Figs. 1(a) and 1(b), respectively. 
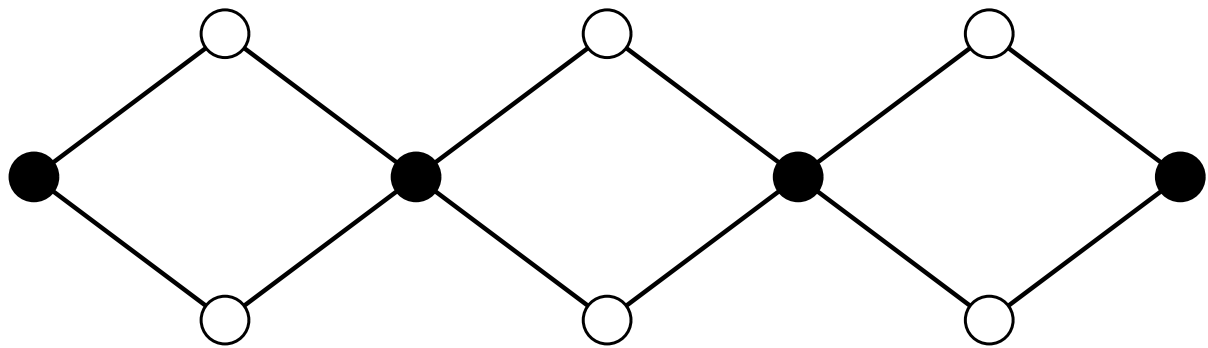
\begin{figure}[h]
 \vspace*{2cm}
 \caption{The diamond chain (a) and 
 the alternately-attached leg chain (b).}
\end{figure}
\noindent
The PD chain has been studied as a possible model for 
an experiment of polymer chain ferromagnetism
by Mac\^edo {\it et al.}.~\cite{MDSCFM1} 

 The band structures of the PD and PAAL chains 
are seen in Figs. 2(a) and 2(b), respectively.
\begin{figure}[h]
 \vspace*{2cm}
 \caption{The energy bands of the PD chain (a) and 
 the PAAL chain (b).}
\end{figure}
Here, ``band'' means a dispersion relation of the single-electron 
state. The PD and PAAL chains have a flat-band, 
namely, a completely-degenerate dispersion relation. 
Although the band cannot be defined on the 
OD chain because of the absence of periodicity, an equivalent 
degeneracy exists in the single-electron energy spectrum of 
the OD chain. 

 Nagaoka's theorem holds only in the PD chain since this model  
satisfies the connectivity condition for $N_{\rm h}=1$. 
The connectivity condition means that the Hamiltonian matrix 
is irreducible, namely, in a physical picture, all of states in 
a subspace with a fixed magnetization can be produced 
from any state by permitted 
motions of electrons. In the PAAL and OD chains, the connectivity  
is not satisfied for $N_{\rm h}=1$. 

 The PD and OD chains have a common characteristic that they have 
a local-loop structure. Here, the local loop means a loop with three 
or four sites (a triangle or a square, respectively). 
It should be noted 
that the local loop itself always satisfies the connectivity condition 
for $N_{\rm h}=1$ while loops with more than four sites do not 
satisfy the condition. 

 At half-filling, it is also noted that all these models 
have the ferrimagnetic ground state because of the difference of 
the numbers of sites in the two sublattices, which has been rigorously 
proven by Lieb.~\cite{LEH1,SSQ1}

\section{Role of the Flat Band}

 In the following two sections, we present numerical results obtained 
by the numerical diagonalization technique (Lanczos method). 
We consider, hereafter, the case that $U=\infty$ as a limit of 
strong correlation. 

 First of all, we calculate the total spin of the ground state in 
the 12 sites system of the PD chain and the PAAL chain with 
the infinitely-large $U$ for $N_{\rm e}=7,8,9,10$ and $11$. 
Results are shown in Table I.  
\begin{table}[h]
 \vspace*{2.5cm}
 \caption{The total spin of the 12 sites PD and PAAL chains and their
 anti-periodic versions.}
\end{table}
In the PD chain, the ground states for high (more than $1/3$) 
electron fillings 
are the saturated ferromagnetic state for odd numbers of 
electrons and the spiral state for even numbers. 
The spin-spin correlation function, 
as an example, for $N_{\rm e}=10$ is plotted in Fig. 3. 
\begin{figure}[h]
 \vspace{2cm}
 \caption{The spin-spin correlation function for $N_{\rm e}=10$.}
\end{figure}
In fact, we see that the spins within a half of system size are 
aligned ferromagnetically. 
We note here that for $N_{\rm e}\le 6$, the ground state is always 
singlet state and not the spiral state. 
The dependence of $S_{\rm tot}$ on $N_{\rm e}$ is shown in Fig. 4. 
\begin{figure}[h]
 \vspace*{2cm}
 \caption{The change of total spins of the 12 sites PD chain in high 
 electron fillings.}
\end{figure}
\noindent
These results show that the ground states for these fillings are 
the extended ferromagnetic state. For the anti-periodic chain, 
the even-odd property of electron numbers reverses. 
This is the same as the spiral state in the other 
models.~\cite{KA1,WM2,KA2}
We would stress that ferromagnetism, but not ferrimagnetism, appears 
in the PD chain. This observation is not intuitively understandable 
from the naive picture of the spin alignment due to the flat band 
for a sufficiently-small $U$. 
This shows that the extended ferromagnetism is realized 
by a true many-body effect that electrons in all the bands correlate 
each other. 
We also study the 18 sites PD chain. Although the calculation 
is limited in subspaces with large magnetization, the results are 
consistent with the results in the 12 sites chain.~\cite{note1} 

 On the other hand, the ground state of the PAAL chain is (disconnected) 
paramagnetic or singlet state, neither ferromagnetic nor ferrimagnetic. 
Moreover, the singlet state is not the spiral state. 
These results do not change for the anti-periodic boundary condition. 
These are also summarized in Table I. 
The results show that the spins even in the flat band do not align, 
at least, in the strongly interacting case except for the half-filled. 
The same assertion has been pointed out in the other model 
by Arita {\it et al.}.~\cite{AKAYTWIOH1} 
These results also deny the following naive picture: 
when $U$ is large, only up (or down) spins are filled in the 
flat band and thus a ferrimagnetism may appear $\rho_{\rm e}\le 1/3$. 

 These results for the PD chain and the PAAL chain show that 
the existence of the flat band is not a sufficient condition 
for the appearance of the extended ferromagnetic state
while the condition for Nagaoka's theorem to hold is important 
for it.

\section{Ferromagnetism without Nagaoka's Theorem}

 Next, we further study a model where Nagaoka's ferromagnetism dose 
not appear but the extended ferromagnetic state appears for 
$N_{\rm h}\ge 2$. As an example of such a model, 
we investigate the OD chain which dose not satisfy the 
connectivity condition at one hole. 
We calculate the OD chains with 10 and 16 sites. 
The results are summarized in Table II. 
\begin{table}[h]
 \vspace{2.5cm}
 \caption{The total spin of the ground state for the OD chain.}
\end{table}
For two holes in 10 sites chain and $2\le N_{\rm h} \le 4$ in 16 
sites chain, 
the ground state is found to be the saturated ferromagnetic state. 
We note here that this is the first result about the appearance of 
the saturated ferromagnetism in such a wide range 
of fillings on the system to which Nagaoka's theorem dose not apply
as far as we know. 
It is reasonable that the spiral state does not exist in the OD chain 
since the spiral state only exists for periodic systems as seen 
in other models.~\cite{KA2,KK1,SUT1,AKKA1} 

 The result of the OD chain shows that the local-loop structure 
plays an important role to realize the extended ferromagnetic state. 
Noting that when $N_{\rm h}\ge 2$, the connectivity condition is 
satisfied even in the OD chain, 
we can obtain the following picture about the 
realization mechanism of the extended ferromagnetic state. 
The local loop creates a local ferromagnetic moment by local hopping 
since the local loop itself satisfies the connectivity condition 
as stated in \S 2. The global ferromagnetic moment, that is, the 
extended ferromagnetism is created by global hopping of electrons 
which is permitted if the connectivity condition is satisfied. 
The fact that the ground state is singlet (not the spiral state) for 
$\rho_{\rm e}\le 1/3$ suggests that the electron hopping is independent 
in each local loop when the electron filling is sufficiently low. 
Thus, we confirm that the high electron filling condition 
is another important factor to realize the extended ferromagnetic state. 
In the above picture, we can understand why the ground state of 
the PAAL chain is not ferromagnetic in spite that the PAAL chain 
also satisfies the connectivity condition when $N_{\rm h}\ge 2$. 
Because, the PAAL chain, in the first place, cannot create 
the local moment for the absence of the local-loop structure. 
This picture is consistent with the observations 
ferromagnetic state in other Hubbard-type models, for example, 
the existence of the spiral state for two holes in the single-band 
Hubbard model on the square lattice 
and so on.~\cite{KA1,WM2,DN1,KM1,AKKA1}

\section{Summary and Discussions}

 We have investigated three models, the periodic diamond (PD) chain, 
the periodic alternately-attached leg (PAAL) chain and the open diamond 
(OD) chain. These models belong to different categories classified by 
the following three characteristics: the flat-band , 
the local-loop structure and Nagaoka's theorem. Although the 
flat band (or the equivalent degeneracy of eigenvalues) exists 
in all of the chains, the local-loop structure exists in the PD and 
OD chains and Nagaoka's theorem holds only in the PD chain.

 First, we calculated the PD and PAAL chains by the exact 
diagonalization technique. 
The result for the PD chain is that the ground state for  
the infinitely-large $U$ 
is the extended ferromagnetic state for $1/3<\rho_{\rm e}<1/2$. 
The result for the PAAL chain, which is contrary to the PD chain, 
the spins do not align except $\rho_{\rm e}=1/2$, at least for 
the infinitely-large $U$. 
Comparing the results of the PD and PAAL chains, 
it is concluded that the existence of flat band does not 
necessarily cause ferromagnetism. That is, 
it is shown that the extended ferromagnetic state is not realized 
by the flat-band. 
Moreover, we stress that the extended ferromagnetism is realized 
by a true many-body effect which can not be understood from the 
single-electron picture like the spin-alignment mechanism of flat band 
for a sufficiently-small $U$ case. 

 Next, calculating the OD chain, we found 
that the saturated ferromagnetic state, remarkably, appears in 
the wide range of electron fillings in spite of the absence of 
the condition for Nagaoka's theorem to hold. The result shows that 
the existence of the local-loop 
structure is essentially important for the appearance of 
the extended ferromagnetism. We, therefore, obtain the following 
picture to realize the extended ferromagnetic state. Once the local 
ferromagnetic moments are generated by electron hopping in a local 
loop, the extended ferromagnetic state are realized by global hopping 
of electrons if the connectivity condition are satisfied and the 
electron filling is sufficiently high. 
This picture shows again that the extended 
ferromagnetism appears due to a purely many-body effect. 

 Here, let us consider the case of half-filling. We point out that
the existence (or absence) of the extended ferromagnetic state 
seems independent on the property at half-filling from the present 
results where the ground state at half-filling is commonly 
ferrimagnetic. The result in the 1D Mielke-Tasaki model also supports 
this statement since in the model, the ground state for half-filling 
is singlet while the ground state for $1/4< N_{\rm e}< 1/2$ is 
the extended ferromagnetic state.~\cite{WM1,WM2} 

 We note, at last, that the above picture to realize the extended 
ferromagnetic state can be easily generalized to 
non-Hubbard-type models in which the extended ferromagnetic state 
appears: the double-exchange model,~\cite{KK1} 
the Kondo-lattice model,~\cite{SUT1} two-band model,~\cite{KA2} 
and the ferromagnetic $t$-$J$ model.~\cite{WM3} 
In these models, the spin-exchange interaction,  
instead of the local-loop structure, generates the local ferromagnetic 
moment. We think that the picture obtained in the present paper 
also give the essential insight for those models in the above sense.

\section*{Acknowledgements}

 We would like to thank Akinori Tanaka for stimulating discussions and 
useful comments. We are also grateful to Shun-Qing Shen for 
valuable suggestions at the beginning of this study. The present work 
is partially supported by Grant-in-Aid from the Ministry of Education, 
Science and Culture. We also appreciate for the facility of Supercomputer 
Center of Institute for Solid State Physics, University of Tokyo.


\begin{thebibliography}{99}
\bibitem{TH0} H. Tasaki: 
Prog. Theor. Phys. {\bf 99} (1998) 489.
\bibitem{PDBDF1} P. Pieri, S. Daul, D. Baeriswyl, M. Dzierzawa 
and P. Fazekas: Phys. Rev. B {\bf 45} (1996) 9250.
\bibitem{GMC1} M. C. Gutzwiller: 
Phys. Rev. Lett. {\bf 10} (1963) 275; 
Phys. Rev. {\bf 137A} (1965) 1726.
\bibitem{HJ1} J. Hubbard: 
Proc. R. Soc. (London), Ser. A {\bf 276} (1963) 159; 
{\it ibid.} A {\bf 277} (1964) 237; 
{\it ibid.} A {\bf 281} (1964) 401. 
\bibitem{KJ1} J. Kanamori: 
Prog. Theor. Phys. {\bf 30} (1963) 275.
\bibitem{NY1} Y. Nagaoka: 
Solid State Commun. {\bf 3} (1965) 409; 
Phys. Rev. {\bf 147} (1966) 392.
\bibitem{TDJ1} D. J. Thouless: 
Proc. R. Soc. (London) {\bf 86} (1965) 893.
\bibitem{TH1} H. Tasaki: 
Phys. Rev. B {\bf 40} (1989) 9192.
\bibitem{MA1} A. Mielke: 
J. Phys. A: Math. Gen. {\bf 24} (1991) L73; 
{\it ibid.} A: Math. Gen. {\bf 24} (1991) 3311; 
{\it ibid.} A: Math. Gen. {\bf 25} (1992) 4335; 
Phys. Lett. A {\bf 174} (1993) 443.
\bibitem{TH2} H. Tasaki: 
Phys. Rev. Lett. {\bf 69} (1992) 1608.
\bibitem{MT1} A. Mielke and H. Tasaki: 
Commun. Math. Phys. {\bf 158} (1993) 341.
\bibitem{TH3} H. Tasaki: 
Phys. Rev. Lett. {\bf 73} (1994) 1158; 
J. Stat. Phys. {\bf 84} (1996) 535; 
Phys. Rev. Lett. {\bf 75} (1995) 4678.
\bibitem{TM1} M. Takahashi: 
J. Phys. Soc. Jpn. {\bf 51} (1982) 3475.
\bibitem{DW1} B. Do\c{c}ot and X. G. Wen: 
Phys. Rev. B {\bf 40} (1989) 2719.
\bibitem{RY1} J. A. Riera and A. P. Young: 
Phys. Rev. B {\bf 40} (1989) 5285.
\bibitem{FRDSR1} Y. Fang, A. E. Ruckenstein, E. Dagotto and 
E. Schmitt-Rink: 
Phys. Rev. B {\bf 40} (1989) 7406.
\bibitem{KA1} K. Kusakabe and H. Aoki: 
Phys. Rev. B {\bf 52} (1995) R8684.
\bibitem{WM1} Y. Watanabe and S. Miyashita: 
J. Phys. Soc. Jpn. {\bf 66} (1997) 2123.
\bibitem{WM2} Y. Watanabe and S. Miyashita: 
J. Phys. Soc. Jpn. {\bf 66} (1997) 3981.
\bibitem{MDSCFM1} A. M. S. Mac\^edo, M. C. dos Santos, 
M. D. Coutinho-Filho and C. A. Mac\^edo: 
Phys. Rev. Lett. {\bf 74} (1995) 1851. 
\bibitem{LEH1} E. H. Lieb: 
Phys. Rev. Lett. {\bf 62} (1989) 1201; 
{\it ibid.} {\bf 62} (1989) 1927(E).
\bibitem{SSQ1} For comprehensive review, see S.-Q. Shen: 
Int. J. Mod. Phys. B {\bf 12} (1998) 709.
\bibitem{KA2} K. Kusakabe and H. Aoki: 
Physica B {\bf 194-196} (1994) 217.
\bibitem{note1} Within the result we can calculate, we regard 
the case that the lowest eigenvalues in all the subspaces are 
identical as the saturated ferromagnetic ground state, 
and the case that the lowest eigenvalue in the subspace with 
the lowest magnetization is unique as the singlet state. Moreover, 
we distinguish the spiral state and the ordinary singlet state 
by the difference of the results the anti-periodic boundary 
condition. 
\bibitem{AKAYTWIOH1} R. Arita, K. Kuroki, H. Aoki, A. Yamaji, 
M. Tsukada, S. Watanabe, M. Ichuimura, T Onogi and H. Hashizume: 
Phys. Rev. B. {\bf 57} (1998) R6854.
\bibitem{KK1} K. Kubo: 
J. Phys. Soc. Jpn. {\bf 51} (1982) 782.
\bibitem{SUT1} M. Sigrist, K. Ueda and H Tsunetsgu: 
Phys. Rev. B {\bf 46} (1992) 175.
\bibitem{AKKA1} R. Arita, K. Kusakabe, K. Kuroki and H. Aoki: 
Phys. Rev. B. {\bf 58} (1998) R11833.
\bibitem{DN1} S. Daul and R. M. Noack: 
Z. Phys. B {\bf 103} (1997) 293; Phys. Rev. B {\bf 58} (1998) 2635. 
\bibitem{KM1} M. Kohno: 
Phys. Rev. B {\bf 56} (1997) 15015.
\bibitem{WM3} Y. Watanabe and S. Miyashita: 
{\it in preparation.}
\end{thebibliography}
\end{document}